\documentclass[epj]{svjour}
\usepackage{epsfig}
\usepackage{subfigure}
\usepackage{amsmath}
\usepackage{amssymb}
\usepackage{array}
\usepackage{pifont}
\usepackage{color}
\newcommand{\rmi}{\textrm{i}}
\newcommand{\rme}{\textrm{e}}
\newcommand{\rmd}{\textrm{d}}
\begin{document}
    \title{Simple model for a quantum wire III. Transmission of finite samples with correlated disorder}%
    \titlerunning{Transmission of finite samples with correlated disorder}
    \date{Received ??? \\ Published online ???}
    \author{J. M. Cerver\'o \and A. Rodr\'{\i}guez}%
    \institute{F\'{\i}sica Te\'orica, Facultad de %
        Ciencias, Universidad de Salamanca, 37008 Salamanca, Spain}%
    \mail{cervero@usal.es}%
    \abstract{The effect of a continuous model of correlations upon
    one-dimensional finite disordered quantum wires modeled by an array of
    delta-potentials, is analyzed. Although the model proposed is not able to
    include new truly extended states in the spectrum, the transport properties
    of a finite sample are noticeably improved due to the existence of states
    whose localization length is larger than the system size. This
    enhancement of transmission is maximized for relatively short chains.%
    }%
\maketitle
\section*{Introduction}
\label{sec:intro}
    In a previous series of papers \cite{QWI}\cite{QWII}, the authors
have analyzed in detail  a simple model describing the main features shown by a
one-dimensional quantum wire. The potential consists of an array of delta 
potentials, a pattern which has been extensively considered (and indeed it is
nowadays) in the literature \cite{deltas}. In Ref. \cite{QWI} the band structure was
fully analytically solved when the structure is periodically
arranged and the density of states together with the localization properties were
described in the random case, for which several novel features such as the
fractal structure of the DOS were reported. In Ref. \cite{QWII} the random 
model was extended to include statistically correlated disorder in a very
natural manner. The effects of the correlations upon the properties of the
system in the thermodynamic limit were studied, however the question of
whether that type of correlations changes or not the transport properties of real
finite structures was left open, and this is the subject of the present work.
The presence of a correlated disorder in a one-dimensional random system can strongly
change its physical properties, by including new resonant extended states
in the case of short-range correlations \cite{RD}, or with the emergence of
mobility edges for the carriers when long-range correlations appear
\cite{long}. The importance of these correlation phenomena has also
been established for two-dimensional structures \cite{hilke}.

The paper is organized as follows. In Section \ref{sec:review}, we briefly 
review the model focusing on the description of the binary disordered
chain and the techniques used to analyze the transport
properties. In Section \ref{sec:results} a large amount of results is
presented together with a discussion about the effects observed, to close
finally with a section of Conclusions.

\section{Review of the model}
\label{sec:review}
Let us briefly review the basic features of the 1D model proposed. For
a detailed description see references \cite{QWI}\cite{QWII}. The wire is
modeled by a linear array of equally spaced delta potentials with different couplings
following a random sequence. In the completely random case, the properties of the system are then
determined by the couplings $\left(a/a_i\right)$ of the species composing the chain and their
concentrations $\{c_i\}$. The density of states and the localization of the electrons
can be studied in the thermodynamic limit by making use of the functional
equation formalism. It is also possible to introduce short-range
correlations in the structure, modifying the probability of different
atomic clusters to appear in the wire sequence. This can be done by
considering an additional set of probabilities $\{p_{ij}\}$ obeying certain
equations, where $p_{ij}$
means the probability for an $i$-atom to be followed or preceded by a
$j$-atom. Thus the frequency of appearance of binary atomic clusters 
can be altered by this quantities. The probability of finding at any
position the couple $-ij-$($-ji-$) would be $c_i p_{ij}$ or equivalently $c_j p_{ji}$.
Then in the thermodynamic limit the physical properties of such a system
will depend upon the couplings of the species, the concentrations, and the
probabilities $\{p_{ij}\}$.  
This correlated model naturally includes the situation when the
disorder in the wire is completely random, that is just defined
 by the values $p_{ij}=c_j$.

\begin{figure}
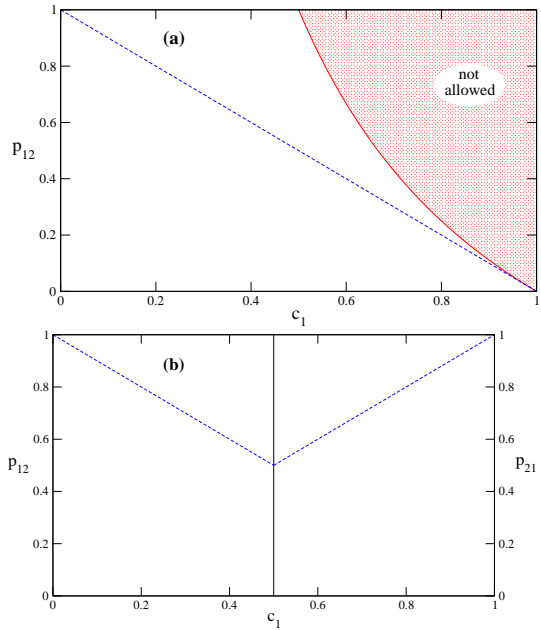

    \centering
    \epsfig{file=fig1.eps,width=.8\columnwidth}
    \epsfig{file=fig2.eps,width=.8\columnwidth}
    \label{fig:space}
    \caption{Correlation space for 2 species as a function of the
concentration. (a) $p_{12}$ vs. $c_1$, (b) optimal representation: if
$c_1\leq0.5$ then $p_{12}$ vs. $c_1$, if $c_1>0.5$ then $p_{21}$ vs. $c_1$.
The blue dashed line corresponds to the completely random configurations.}
\end{figure}

In this work only binary chains are considered, so let us study in detailed
the correlation scheme for this case. Our wire will be determined by
 one of the  concentrations $\{c_1,c_2\}$ and one the probabilities
$\{p_{11},p_{12},p_{21},p_{22}\}$, that satisfy the relations 
$p_{11}+p_{12}=p_{22}+p_{21}=1$ and $c_1 p_{12}=c_2 p_{21}$.
One usually takes as configuration parameters $c_1\leq1$ and $p_{12}\leq\min\{1,c_2/c_1\}$.
 The allowed configuration space with these parameters is shown
in Figure \ref{fig:space}(a). However one can optimize the representation
of this space by choosing the parameters $\{c_1,p_{12}\}$ when $c_1\leq0.5$
and $\{c_1,p_{21}\}$ when $c_1>0.5$, so that the configuration
space is expanded and the spatial points can be better differentiated, 
as shown in Figure \ref{fig:space}(b).
Therefore, for a given concentration different values for $p_{12}$($p_{21}$) can be chosen, and only one of them corresponds to
the completely random chain. When the configuration of the binary chain
 lies on the dashed lines of Figure \ref{fig:space}, we have a completely
random chain whereas if the configuration lies anywhere else we have a
correlated chain.

The main aim of this work is to elucidate whether or not this type of
short-range correlations can improve the transport properties of real
finite systems. With this purpose the following techniques are used, in
contrast with the ones used in \cite{QWI}\cite{QWII} for treating  
infinite systems.
\subsection{Transmission matrix formalism}
The time-independent scattering process in one-dimension can be described
using the well-known continuous transfer matrix method \cite{harrison}. The
transmission matrix for a delta potential preceded by a zero potential zone
of length $a$ can be easily calculated yielding,
\begin{equation}
    \label{eq:tmdelta}
    \mathbf{M}_j(k)=\begin{pmatrix} \left(1-\frac{\rmi}{ka_j}\right)\rme^{\rmi
ka} & 
    -\frac{\rmi}{ka_j} \rme^{-\rmi ka} \\ \frac{\rmi}{ka_j} \rme^{\rmi ka} & 
    \left(1+\frac{\rmi}{ka_j}\right)\rme^{-\rmi ka} \end{pmatrix}
\end{equation}
where $a_j=\hbar^2/(m\alpha_j)$ means the ``effective range'' of the
$j$th delta, being $\alpha_j$ its coupling. The composition of $N$
potentials can then be considered through the product of 
matrices,
\begin{equation}
     \mathbb{M}=\mathbf{M}_N\ldots\mathbf{M}_2\mathbf{M}_1
\end{equation}
to obtain the global transmission from $T(k)=|\mathbb{M}_{22}|^{-2}$. 
This formalism can be numerically applied to consider large chains but
one finds also that for delta potentials it is possible to write analytical
closed expressions for the scattering amplitudes of a chain composed of $N$
different units \cite{pra}.

Once the transmission of the finite sample has been calculated, the
inverse of the localization lenght of the electronic states can be characterized by the Lyapunov
exponent via the expression \cite{kirkman}\cite{kramer}
\begin{equation}
    \Lambda(k) =-\frac{1}{2N}\log T(k).
    \label{eq:lya}
\end{equation}
\section{Results}
\label{sec:results}
The effects of the correlations upon the system at the thermodynamic limit
were extensively analyzed in Ref. \cite{QWII}. The authors concluded that
the density of states is drastically changed by the effect of
correlations. For a wire with fixed concentrations, the correlations can be
tuned to open or close gaps in the spectrum, and they alter the number of
available states at a certain energy as well as the smooth or irregular
evolution of the DOS. Concerning the spatial extension of the electron wave
functions, the influence of the correlations on the localization properties
was established. An important change on the localization length was
observed for all energies. The value of the Lyapunov exponent could be
greatly decreased for some energies at the expense of an increasing
behaviour in other ranges. However this type of correlations does not cause
the appearance of neither new resonant energies nor mobility edges for the
carriers. The question of whether these correlations might change the transport
properties of a finite system was left open.

Let us have a look at the transmission patterns of finite binary chains for
different configurations of concentrations and correlations. 
\begin{figure}
    \centering
    \epsfig{file=fig3.eps,width=.9\columnwidth}
    \epsfig{file=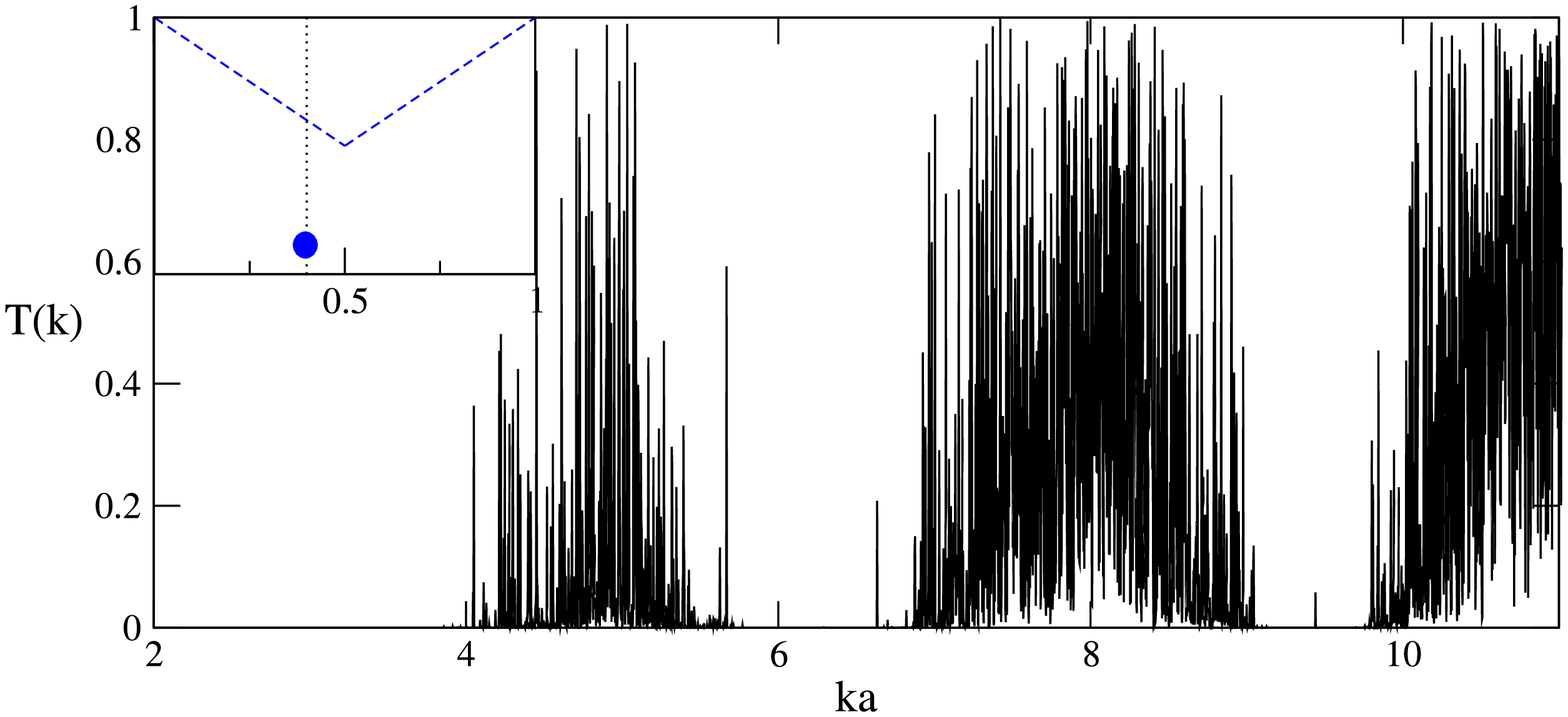,width=.9\columnwidth}
    \epsfig{file=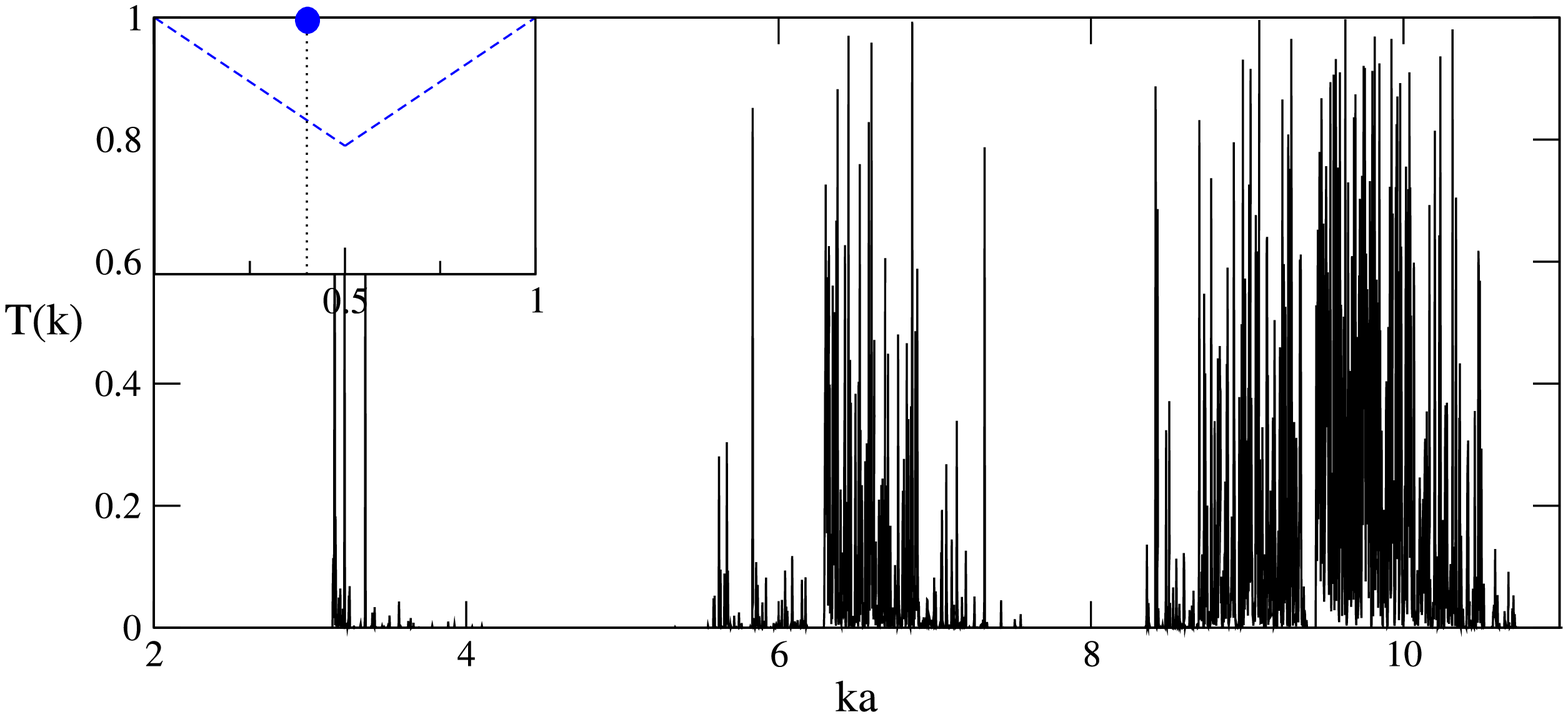,width=.9\columnwidth}
    \caption{Transmission probabilities vs energy for 1000 atoms binary
disordered chains with couplings $(a/a_1)=1,\,(a/a_2)=-1$ and
concentrations $c_1=0.4,\,c_2=0.6$ for different correlation
configurations. From top to bottom $p_{12}=0.6,0.1,1.0$. The circular point inside
the insets mark the configuration on the correlation space. Only one
realization of the disorder has been considered for each case.}
    \label{fig:tran1}
\end{figure}
\begin{figure}
    \centering
    \epsfig{file=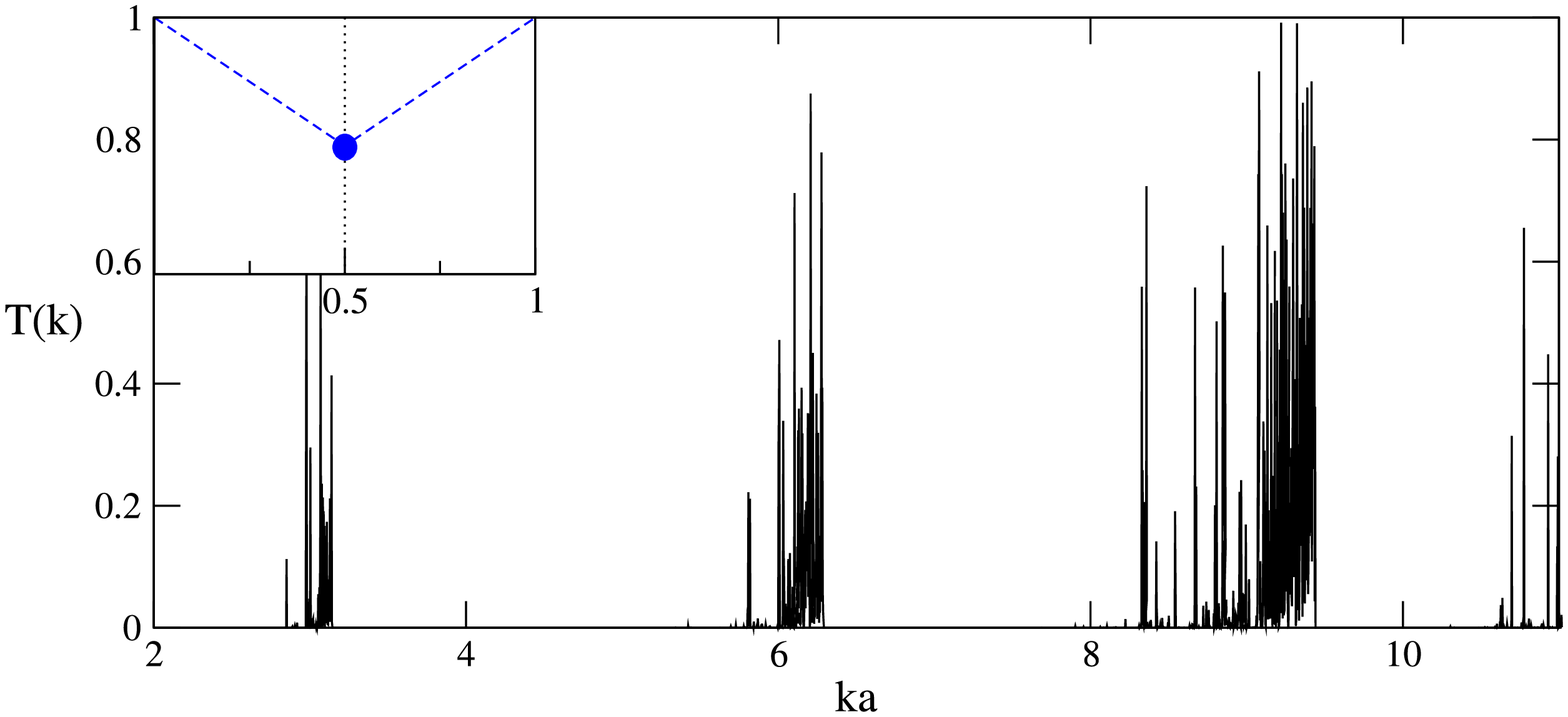,width=.9\columnwidth}
    \epsfig{file=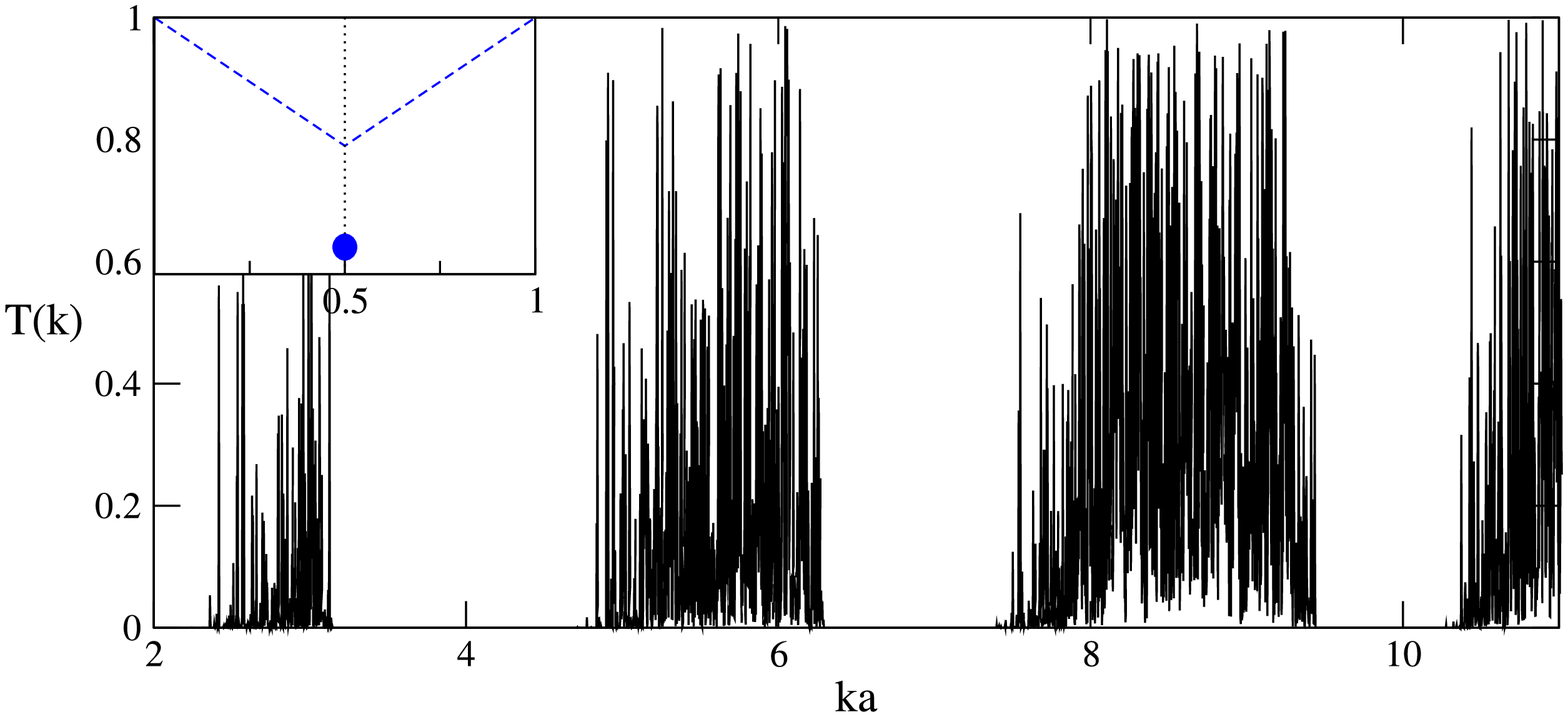,width=.9\columnwidth}
    \epsfig{file=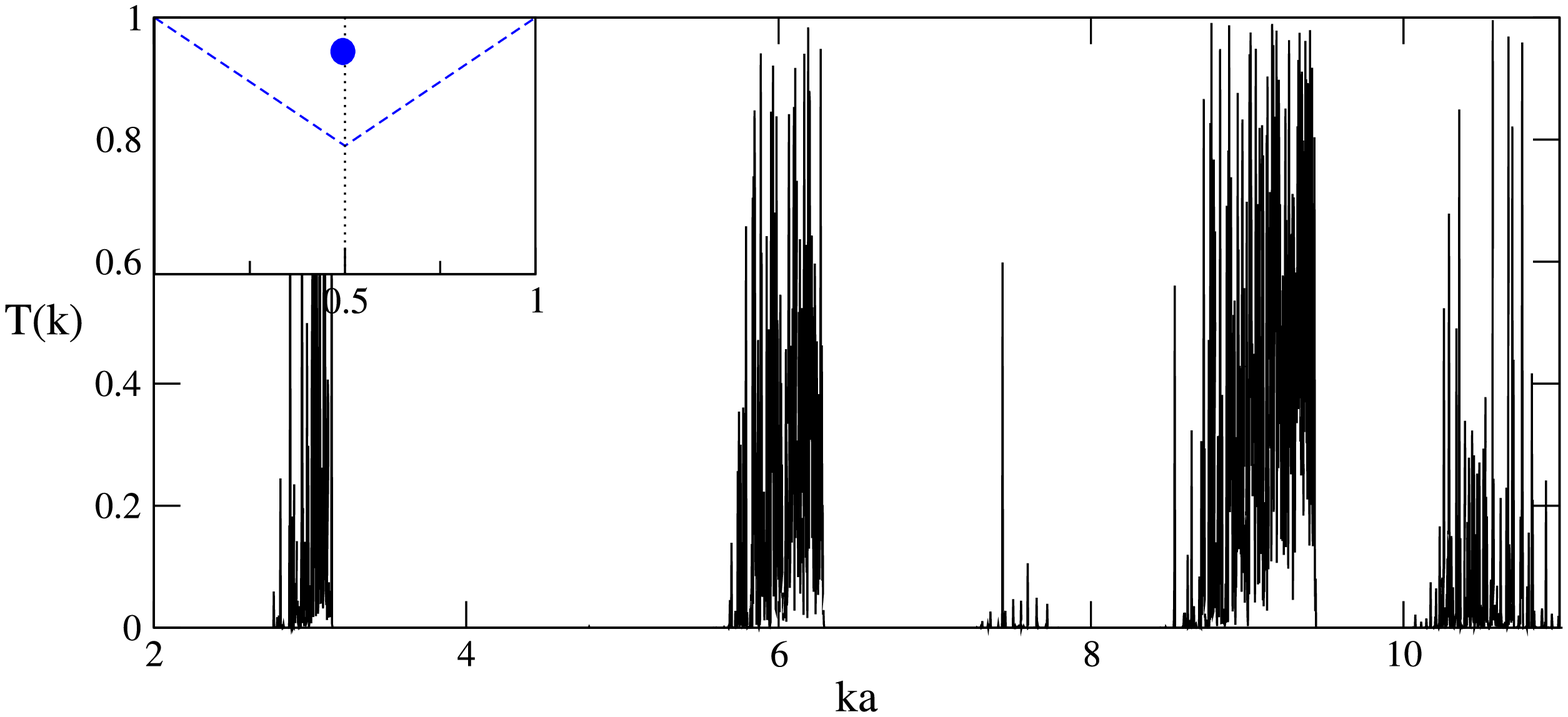,width=.9\columnwidth}
    \caption{Transmission probabilities vs energy for 1000 atoms binary
disordered chains with couplings $(a/a_1)=2,\,(a/a_2)=4$ and
concentrations $c_1=c_2=0.5$ for different correlation
configurations. From top to bottom $p_{12}=0.5,0.1,0.85$. The circular point inside
the insets mark the configuration on the correlation space. Only one
realization of the disorder has been considered for each case.}
    \label{fig:tran2}
\end{figure}
In Figures \ref{fig:tran1} and \ref{fig:tran2} the transmission is shown
for several chains composed of 1000 atoms, for different values of the couplings and concentrations. In
these cases the worst transmission corresponds to the completely random
configurations, for which the transmission probability  only
raises near the multiples of $\pi$ due to the well known resonances
of the model at these energies. However as we move away from the completely
random configuration (above or below the dashed line) the transmission is
noticeably improved. Notice that this improvement is not necessary localized 
around the multiples of $\pi$. Although quantitatively this enhancement 
depends on the values of the couplings, qualitatively it seems a generic 
behaviour. In order to check whether this effect can be extended over the
whole correlation space, we characterize each of its points by an efficiency
of transmission defined as,
\begin{equation}
    T_{\textrm{eff}}=\frac{1}{k_2-k_1}\int_{k_1}^{k_2} T(k) \rmd k
    \label{eq:teff}
\end{equation}
which is the area enclosed by the transmission coefficient per energy unit.
This definition depends on the integration interval, but
qualitatively the results will not be affected as long as a reasonable interval is chosen, generally one of the form
$[0,k_2]$. Notice that for very high energies the transmission will
saturate for all configurations, thus the contribution to the integral in
\eqref{eq:teff} will be the same independently of the $c_1,\,p_{12}$ values. 
We are interested in establishing a qualitative comparison of this efficiencies for different correlations. 

\begin{figure}
    \centering
    \epsfig{file=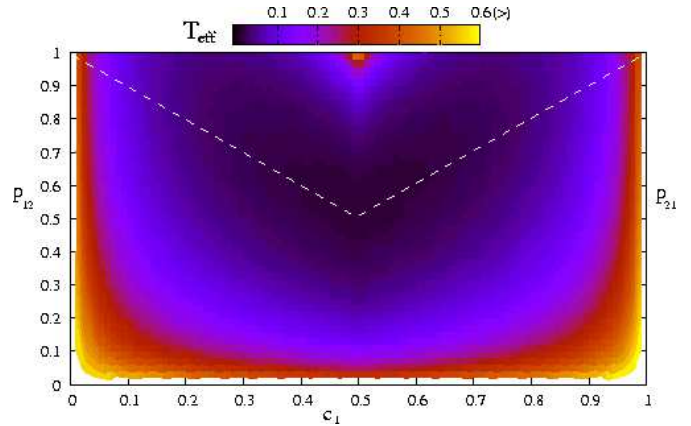,width=\columnwidth}
    \caption{Transmission efficiency for different configurations of a
    binary chain with $1000$ atoms and couplings  $(a/a_1)=1,\,(a/a_2)=-1$. For
    each configuration only one realization of the disorder has been
    considered. The integration interval for $T_\textrm{eff}$ was
    $[0,15]$.}
    \label{fig:map}
\end{figure}

\begin{figure}
    \centering
   \epsfig{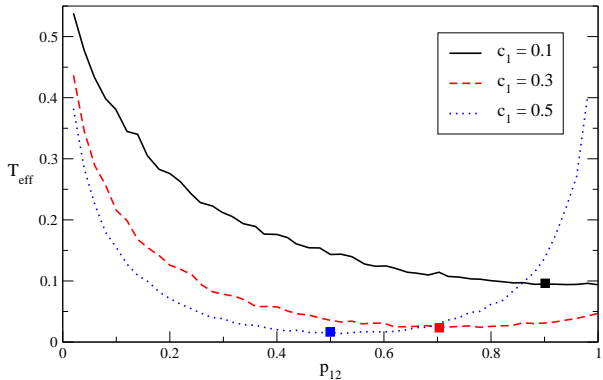}
    \caption{$T_\textrm{eff}$ vs $p_{12}$ for $1000$ atoms binary chains
    with couplings $(a/a_1)=1,\,(a/a_2)=-1$ and different 
    concentrations. The squares on the lines mark the position of the completely
    random configuration.}
    \label{fig:cuts}
\end{figure}

For  certain values of the couplings and a length of $1000$ atoms the evolution of
this transmission efficiency over the configuration space is shown in
Figure \ref{fig:map}.
It is clearly shown that the lowest values for the transmission efficiency
are distributed around the completely random configurations, specially when
the participation of the species is homogenized ($c_1\sim 0.5$). High
efficiencies can be observed for low and high concentrations of one of the
species (and therefore approaching a pure chain) and around the point
$\{c_1=0.5,p_{12}=1.0\}$ which corresponds to the periodic binary
chain. Nevertheless by looking at the evolution of $T_\textrm{eff}$ as a
function of $p_{12}$ for a fixed concentration (Figure \ref{fig:cuts}) we
conclude that the minimum efficiency is reached near the completely random
configuration and the correlated situations show noticeably higher values.

\begin{figure}
    \centering
    \epsfig{file=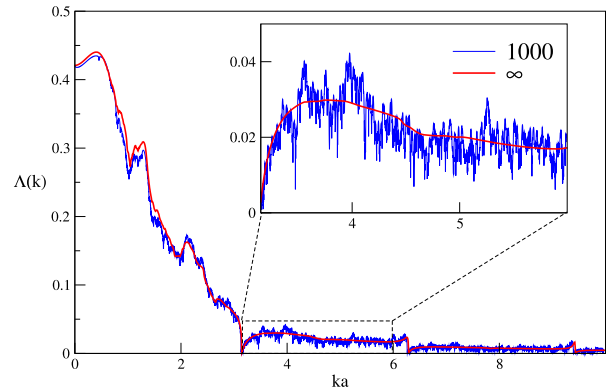,width=.9\columnwidth}
    \caption{Lyapunov exponent vs energy for a binary chain with parameters
    $(a/a_1)=1,\,(a/a_2)=-1$, $c_1=0.4$, $p_{12}=0.6$. The blue line
    corresponds to a $1000$ atoms realization and the red line to the
    infinite chain.}
    \label{fig:lyaex}
\end{figure}

\begin{figure}
    \centering
    \epsfig{file=fig12.eps,width=.85\columnwidth}
    \epsfig{file=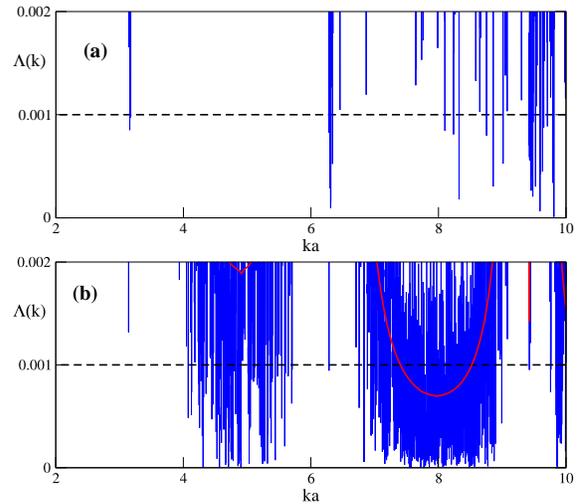,width=.85\columnwidth}
    \caption{Lyapunov exponent vs energy for a $1000$ atoms binary chain
    with couplings $(a/a_1)=1,\,(a/a_2)=-1$ and concentration $c_1=0.4$ for (a)
    completely random configuration $p_{12}=0.6$ and (b) correlated
    configuration $p_{12}=0.1$. The dashed line marks the inverse of the length
    of the sample. The red line shows the Lyapunov exponent for the
    infinite chain.}
    \label{fig:lyazoom}
\end{figure}

Therefore the electronic transmission trough a finite wire is improved by
this type of correlations although truly extended states do not appear in
the system. The reason for the improvement then must be the
existence of states behaving as extended states,
that is their localization length being larger that the system size. Let us
analyze the behaviour of the Lyapunov exponent. In Figure \ref{fig:lyaex}
 the Lyapunov exponent as a function of the energy  is shown for a random chain. We can see
a very good agreement between the thermodynamic limit and the finite
realization of the disorder, that shows a characteristic fluctuating behaviour around the
values of the former one. These fluctuations are responsible for the enhancement of
transmission. A fine observation of the Lyapunov
exponent, in Figure \ref{fig:lyazoom}, reveals that for a chain with fixed
concentrations the number of states whose localization
length exceeds the sample length increases dramatically in a correlated
configuration with respect to the completely random situation. The correlations
induce a decrease of the limiting distribution of
the Lyapunov exponent in certain energy ranges, so that for a finite system the fluctuations of this
quantity around its mean value make the appearance of such states possible.
Let us remark that although the fluctuating pattern is a fingerprint of the
particular realization of the disorder, the amplitude of these oscillations
does only depend upon the length of the system. Therefore the results are
not due to particular bizarre realizations of the disorder for a given
length. The behaviour described can be clearly observed in all realizations of a certain 
configuration. However if one takes averages over several realizations, the
fluctuating behaviour of the Lyapunov is killed as one approaches the
thermodynamic limit. Let us make it clear that  the averaging procedure is
intended to produce the localization properties of the system in that limit.
\begin{figure}
    \centering
    \epsfig{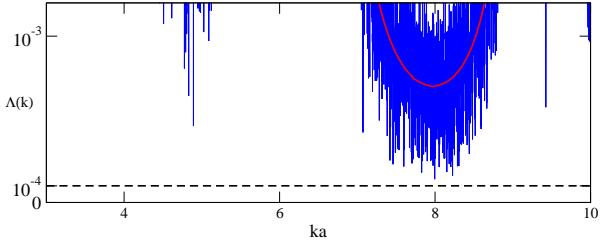}
    \caption{Lyapunov exponent vs energy for a binary chain
    with couplings $(a/a_1)=1,\,(a/a_2)=-1$ in a correlated configuration
    $c_1=0.4$, $p_{12}=0.1$ for $10^4$ atoms. The dashed line marks the inverse of the length
    of the sample. The red line corresponds to the infinite chain. To be
    compared with figure \ref{fig:lyazoom}(b).}
    \label{fig:lyastep}
\end{figure}

\begin{figure}
    \centering
    \epsfig{file=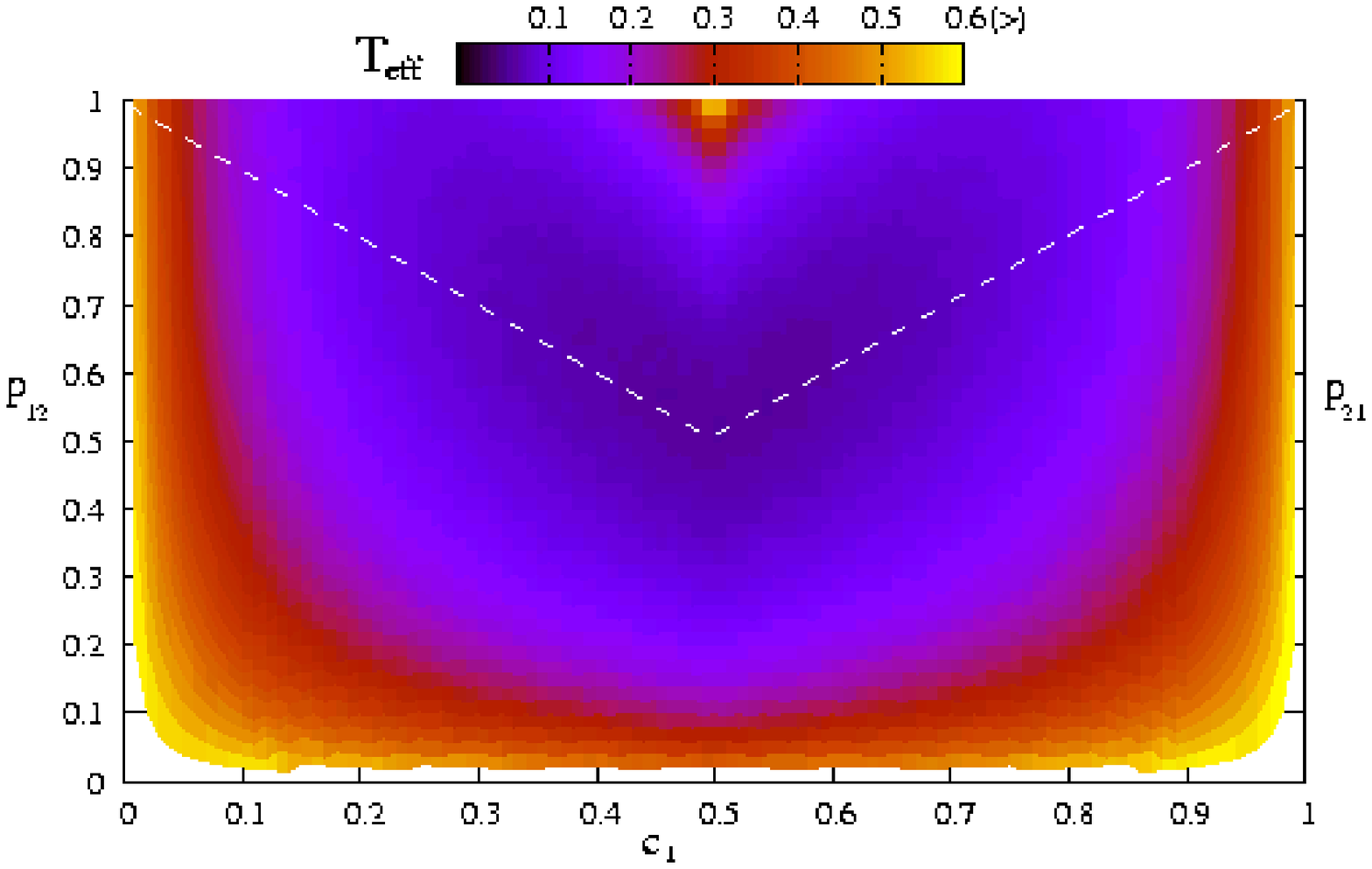,width=\columnwidth}
    \epsfig{file=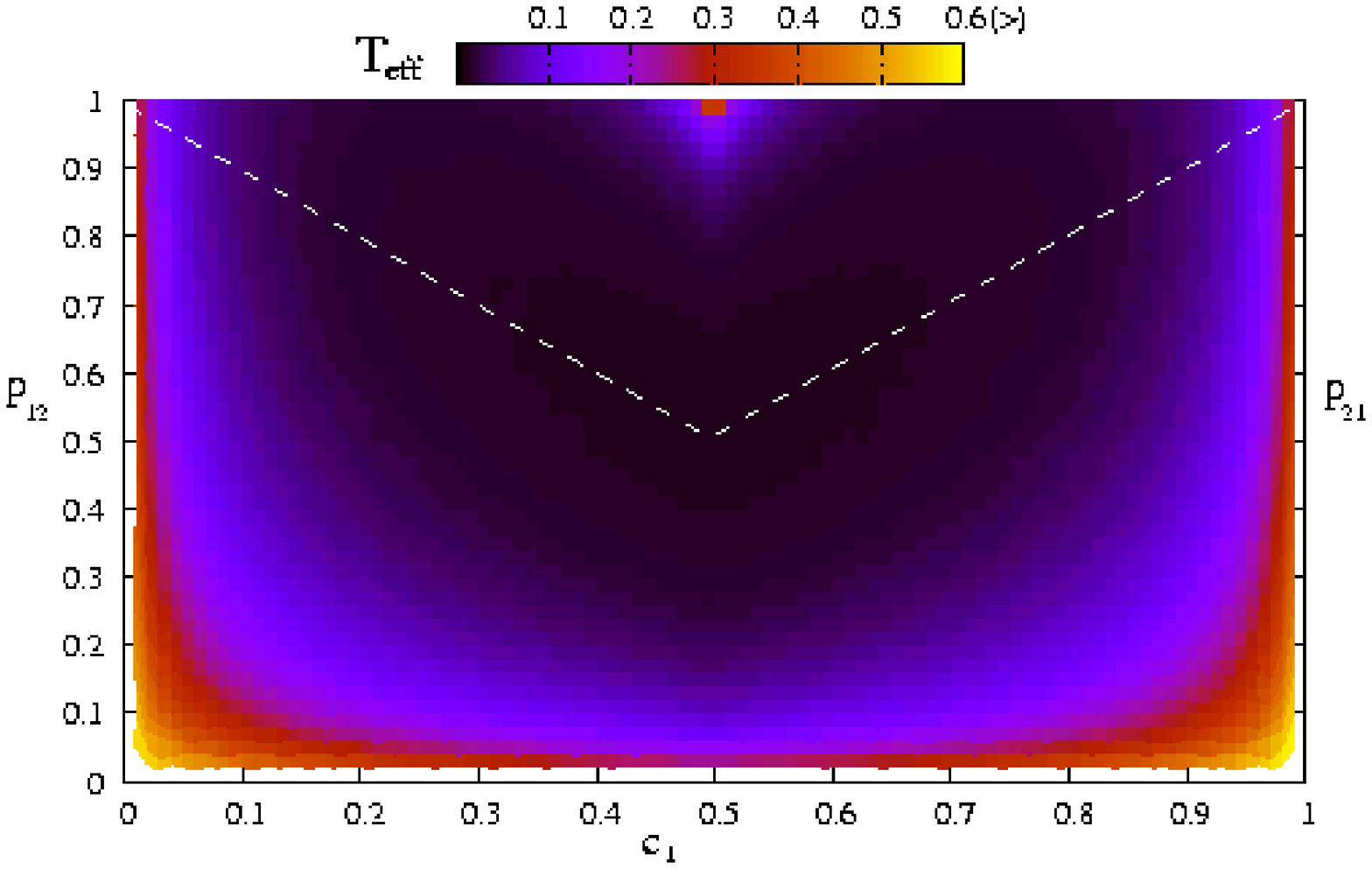,width=\columnwidth}
    \caption{Transmission efficiency over configuration space for a binary
    chain with couplings $(a/a_1)=1,\,(a/a_2)=-1$ for different lengths:
    $L=500$ (top) and $L=2000$ (bottom).}
    \label{fig:mapevol}
\end{figure}

As expected for a model of short-range correlations, all the 
effects disappear unavoidably in the thermodynamic limit. Thus as the
length of the chain grows the fluctuations of the Lyapunov exponent
decrease and the localized character of the electronic states
naturally manifests itself for all energies (Figure \ref{fig:lyastep}). The lost of
the enhancement of transmission can also be shown as a function of the evolution of
$T_{\textrm{eff}}$ over the configuration space for different
lengths. The higher the number of atoms the more the black zones spread 
from the completely random lines (Figure \ref{fig:mapevol}). 
\begin{figure}
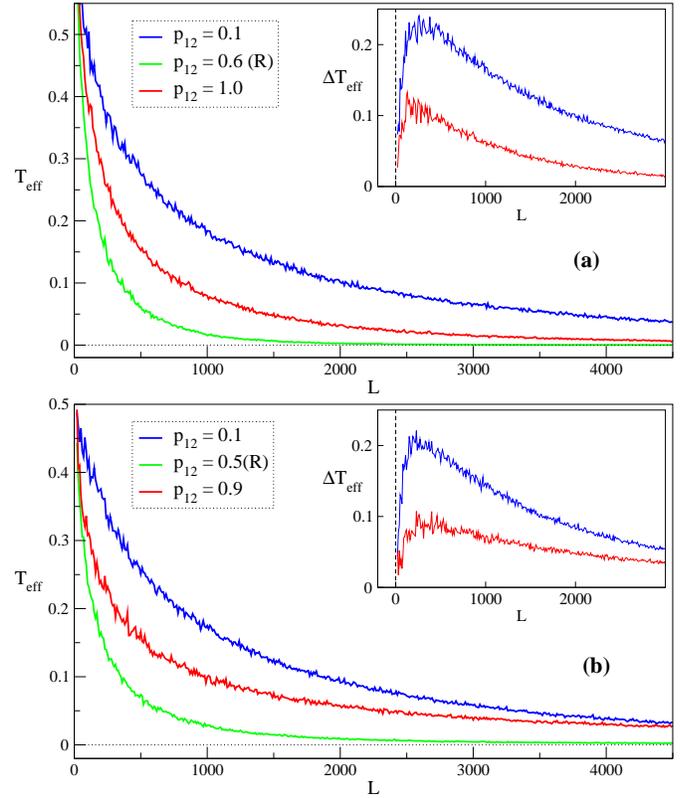

    \centering
    \epsfig{file=fig17.eps,width=\columnwidth}
    \epsfig{file=fig18.eps,width=\columnwidth}
    \caption{Transmission efficiency vs length for different
    configurations of a $1000$ atoms binary chain with parameters (a) $(a/a_1)=1$,
    $(a/a_2)=-1$, $c_1=0.4$ and (b) $(a/a_1)=2$, $(a/a_2)=4$, $c_1=0.5$.
     (R) marks the completely random situation. The inset shows
    the relative differences $\Delta T_\textrm{eff}=T_\textrm{eff}-T_\textrm{eff}(R)$.}
    \label{fig:decay}
\end{figure}
However the decay of the transmission efficiency with the length of the
system depends upon the correlations. In Figure \ref{fig:decay} it can be seen how the
fastest decreasing corresponds to the completely random situation, whereas the
correlated chains show always higher efficiencies for all
lengths. Plotting for different configurations $\Delta
T_\textrm{eff}=T_\textrm{eff}-T_\textrm{eff}(R)$ as a function of the length, where $(R)$ means the
completely random situation, we see
how the effect of the correlations reaches a maximum which is roughly contained in
the region $L\sim 200-500$, apparently independent of the values of
the species couplings.
\section{Conclusions}
To summarize, we have analyzed in detail the effect of a model of
correlations proposed in \cite{QWII}, on finite disordered wires. The
improvement of the transport properties has been established, not only by
looking at the transmission coefficient of particular chains, but
also in the whole correlation space of a binary array through the
transmission efficiency. For fixed concentrations the electronic transport
reaches its minimum intensity near the completely random configuration,
whereas the correlated situations show noticeably higher transmission
efficiencies. The enhancement of the transport properties is due to the
appearance of states with a localization length larger that the system
size that effectively behave as extended states. As the length of the system
grows the effect of these short-range correlations disappears, and the
transmission decreases. The fastest decay corresponds to the completely
random situation. The effect of the correlations as a function of the
length reaches a maximum for relatively short chains $L\sim 200-500$.  

We believe that the behaviour described is essentially independent of the
potential model and that the same effects could be observed for other
models such as the tight-binding scheme or for square barriers, as well as
for the case where more species are included in the wire.
Let us finally remark that although the correlation model considered is not
able to include any new truly extended state in the spectrum, its effects
upon the transport of real finite samples are absolutely non-negligible and
 they may be significant in certain experimental devices such as for
example superlattices, which have already been used to observe
the effect of other models of correlations \cite{bellani}.

\begin{acknowledgement}
    We acknowledge with thanks the support provided by the Research in
Science and Technology Agency of the Spanish Goverment (DGICYT)
under contract BFM2002-02609.
\end{acknowledgement}


\end{document}